\documentclass[twocolumn]{aa}
\usepackage{graphicx}

\begin{document}

\title{The double radio source 3C343.1:  A galaxy-QSO pair with very different
redshifts}

\author{H. Arp
\inst{1}
\and E. M. Burbidge
\inst{2}
\and G. Burbidge
\inst{2}}

\offprints {H. Arp}

\institute{Max-Planck-Institut f\"ur Astrophysik, Karl Schwarzschild-Str.1,
  Postfach 1317, D-85741 Garching, Germany\\
   \email{arp@mpa-garching.mpg.de}
\and
Center for Astrophysics and Space Sciences
0424, University of California, San Diego, CA 92093-0424, USA\\
\email{gburbidge@ucsd.edu}}

\date{Received}

\abstract{
The strong radio source 3C343.1 consists of a galaxy and a QSO
separated by no more than about $0.25^{\prime\prime}$. The chance of this being 
an accidental superposition is conservatively $\sim1\times 10^{-8}$. The 
$z=0.344$ galaxy is connected to the $z=0.750$ QSO by a radio bridge.  
The numerical relation between the two redshifts is that predicted from
previous associations. This pair is an extreme example of many similar physical 
associations of QSOs and galaxies with very different redshifts.
}

\titlerunning{galaxy - QSO pair}

\maketitle

\section{Introduction}

Over the years many cases of QSOs associated with
active galaxies with much smaller redshifts have been discovered.
The papers generally show evidence involving radio emitting QSOs
and bright galaxies, X-ray-emitting QSOs and active galaxies, and
pairs with optical, radio and X-ray connections (Burbidge et al. 1971; 
Pietsch et al. 1994; Burbidge 1995, 1997, 1999; Radecke 1997; Arp 1967, 
1996, 1997,1999, 2003; Arp et al. 1990, 2002).

Two of the most impressive recent examples are the X-ray QSOs
lying very close to the nucleus of NGC 3628 (Arp et al. 2002) and
the discovery of two QSOs in the optical bridge between NGC 7603
and its companion galaxy (L\'{o}pez-Corredoira and Guti\'{e}rrez 2002).

It has also been shown that in many of the cases known, the
redshifts of the QSOs ($z_{\rm q}$) are related to the redshifts of the
parent galaxies ($z_{\rm g}$) by the relation:

   $$(1 + z_{\rm q}) ~= ~(1 + z_{\rm g})(1 + z_{\rm i})(1 + z_{\rm d})$$

\noindent where $z_{\rm i}$ is an intrinsic redshift component, and
$z_{\rm d}$ is a measure of the Doppler shift (either positive or
negative) associated at least partly with the ejection speed of the QSO 
from the nucleus of the galaxy. It is now well established that intrinsic
redshift components follow the law $\Delta$ log (1 + $z_{\rm i}$) =
0.089 with peaks at: 

          $$z_{\rm i} = 0.061, 0.30, 0.60, 0.96, 1.41, 1.96, 2.63 etc.$$

(Karlsson, 1971, 1973, 1977, 1990; Burbidge and Burbidge
1967; Arp et al. 1990; Burbidge and Napier 2001). For nearby bright galaxies from
which most of the bright samples have been taken $z_{\rm g}$ is very
small, and $z_{\rm d}$ is also quite small, so that $z_{\rm q} \simeq z_{\rm i}$.

In this paper we briefly describe the properties of yet another pair
which is much closer in angular separation than most of the cases so far found.

\section{3C 343.1}

\subsection{Optical Properties}
This powerful radio source in the 3C catalogue was first identified and its redshift 
was measured by Spinrad et al. (1977).  They detected a strong emission line 
which they identified as [OII] $\lambda$ 3727 at a redshift $z$ = 0.750.  Thus it 
was classified as a high redshift radio galaxy (Spinrad et al. 1985).
    
This source is one of four high redshift radio galaxies investigated in modern times 
by Tran et al. (1998) with the Low Resolution Imaging Spectrograph (LRIS) on the 
Keck 10 meter telescope. Tran et al. found that the central spectrum showed no 
higher ionization lines other than [OII] $\lambda$ 3727 and [OIII] 
$\lambda\lambda$ 4959, 5007.

However they found that there is a second system present which
gives rise to an absorption line spectrum with high-n Balmer lines
in absorption together with [OII] $\lambda$ 3727 and [OIII]
$\lambda\lambda$ 4959, 5007 emission. The redshift of this system
is $z = 0.344.$  They also showed that [OII] $\lambda$ 3727
emission at the lower redshift can be seen over about
5$^{\prime\prime}$, while the emission line [OII] $\lambda$ 3727
at $z = 0.75$ extends only over the nuclear region. Thus they concluded that two 
separate objects are contributing to the observed spectrum, an underlying active 
galaxy and a high redshift QSO.

\subsection{Radio properties}
In Fig. 1 we reproduce the radio map of 3C 343.1 made by Fanti et
al. (1985).  The source was also mapped at 15 and 22.5 GHz by van
Breugel et al. (1992).  These maps show what is apparently a
classical double-lobed radio source, where usually the galaxy
responsible for the ejected radio emission lies between the two
lobes. In 3C 343.1, however, we apparently have a radio galaxy
emitting along one of its lobes a QSO that is itself a radio
source, the two being of approximately equal intensity.

The relation of the radio emission isophotes to the optical isophotes is
shown in images published by de Vries et al. (1997) where the
Hubble Space Telescope imaging of a number of compact steep-spectrum 
sources is shown. Fig. 2.26 of de Vries et al. shows, at
the top, the HST/WFPC image of 3C 343.1. The optical radiation is
double, aligned (as they measured it) $9^o$ to the E-W direction,
the eastern component being the brighter. Below it they show the
radio image taken with VLBI at 0.6 GHz, and the image at the
bottom of the figure shows the radio contours superposed on the
HST optical image. The correspondence is exact, the radio contours
fit right over the optical image.  It is clear from these maps that the
separation between the two centers is no more than about $0.25^{\prime\prime}.$

\subsection{Identification of components}

We consider the HST image of de Vries et al. (1997), as shown by the top image
in their Fig. 2.26. The western (fainter image) is clearly near the resolution of the
F702W image of the WFPC2 of the Hubble Space Telescope, while the brighter,
eastern component is clearly extended, more in the E-W than in the N-S 
direction. This corresponds to the extension of [OII] $\lambda 3727$ emission
over about $5^{\prime\prime}$ which was detected in the lower redshift spectrum.
Could the images be explained by a single optical structure with a superposed
dust lane across the center? Their brief description gives no indication that this
was considered a possibility. Moreover a dust lane while partially dimming
the optical radiation in the middle panel of Fig. 2.26 can have no effect on the 
radio radiation. The radio emission as shown in the bottom panel of Fig. 2.26, 
with strong central contours in both E and W components, strongly suggests 
two separate radio emitters.

     In summary, regardless of the exact pointing accuracy, there
     is no question that the HST has imaged the radio source 3C343.1
     The source is shown to consist of two optical objects. The Keck
     spectra show two separate spectra, one of z = .34 and one of
     z = .75. The redshifts can be assigned to the appropriate objects
     by their extension and compactness in the optical, spectroscopic
     and radio. But in fact this is only additional confirmation of
     the major point that there are two objects with strong evidence for 
     physical association which have much different redshifts.

     The optical compactness of the object at z = .75 would normally
     qualify it as a QSO. But it could also be called an AGN. Again
     the main result is the association of two much different
     redshifts.

\section{Association between the two components}

In the present case we know that the z = 0.344 object is a galaxy because 
it has narrow emission and absorption lines and is extended on the 
spectrogram of Tran et al. (1992). Moreover, the eastern component of 
the optical image (de Vries et al. 1997) is brighter and more extended than 
the western one, as on would expect a galaxy of that redshift to appear.
The high resolution radio map galaxy with $z$ = 0.344 clearly shows the 
classic bipolar radio ejection coming out in opposite directions from its 
center. As in many of the cases of ejections from galaxies, there is
a QSO at the end of one or both jets (Burbidge 1995; Arp 1996; 
Arp et al. 2002; L\'opez-Corredoira and Guti\'errez 2002).

In the particular case of 3C 343.1 the accurate radio mapping of
the European VLBI Network (Fanti et al. 1985) enables us to
actually see what may be the compression of radio contours as bodies move
through the ambient medium.  Fig. 1 suggests that the QSO is
moving away from the galaxy exactly along the line of the bridge
joining them.  The following material ejected from the galaxy in
this direction, however, is apparently meeting the trail of QSO
material and is compressed by that interaction.

This QSO-galaxy pair is unique because the separation between them
is extremely small.  Both components are seen together in the
spectrum.  The separation between the two centers is
$0.25^{\prime\prime}$ $\simeq 2 kpc$ while the angular size of
the optical emitting region of the lower redshift galaxy is 
$\sim 5^{\prime\prime}$ $\simeq 40 kpc$ $(H_o = 60 km sec^{-1}
Mpc^{-1}$). With such a small separation it is possible that a
radio bridge is being detected before it breaks up and the
galaxy/QSO pair assume the configuration seen in other cases.  It
would be natural to expect the configuration to change rapidly at
first with the radio bridge fading and breaking up as the objects
separated. Such an evolution could explain the rather infrequent
observation of radio bridges between galaxies and their ejecta.

\section{The probabilities}

In view of the fact that the 3C is a complete survey of bright radio sources in the
northern hemisphere it is natural to calculate what are the chances of two of its 
sources accidently falling as close together as those pictured in Fig. 1.

If we say there are 300 radio galaxies in the catalogue, the total area of the sky
within $0.25^{\prime\prime}$ is $\pi x .25^2 x 300 = 4.5\times 10^{-6}$ sq. deg. We
 place randomly one 3C quasar in the  23,000 sq deg. of the Catalog down to 
$Dec. = -5^o.$ The probability that it lies within $0.25^{\prime\prime}$ of any of 
these 300 radio sources is then $4.5\times 10^{-6} / 23,000 = 2 x 10^{-10}$.

There are 50 such 3C quasars so the probability that any lie within 
$0.25^{\prime\prime}$ is:
                  $$50 x 2 x 10^{-10} = 1 x 10^{-8}$$
But this is an overly conservative estimate for two reasons:

1) The radio plasma appears to form a continuous bridge between the galaxy and
the quasar in Fig. 1. If that is accepted there would be no point in computing
probabilities. But if we do not consider the radio material linking them to be a 
physical bridge, we must still estimate the chance that the radio tail from the 
galaxy accidentally points to within better than a few degrees to the quasar and 
similarly from the quasar back to the galaxy. This would give a further 
improbability of $(\pm 2/90)^2 = 5x10^{-4}$. The combined probability of this 
configuration being chance is of the order of: 

                              $$5 x 10^{-12}$$

2) Further double spectra among the 300 may be present but 
unrecognized. There could well be other cases where there are fainter or unidentified 
lines as in the spectra of 3C343.1, one of only four 3C quasars observed with Keck in Tran 
et al. (1998).

Additionally, there is abundant previous evidence for 3C quasars physically
associated with bright active galaxies. In 1971 a paper usually referred to as 
$B^2S^2$ (Burbidge, Burbidge, Strittmatter and Solomon) investigated the 
QSO's in the 3C and 3CR Catalogs. They found that the probability of chance 
association with low redshift galaxies in this entire sample was $<10^{-3}$. This, 
however, was based solely on the criterion of nearness on the sky. In the 
subsequent years some of their closest pairs have shown other evidence for 
association and a number of additional high significance associations have been 
found (Arp 1996; 1998; 2003). If we ask the question what determines the 
probability of an association we can list five empirical criteria: nearness, 
alignment, centering, similarity of ejecta (usually z's or apparent mag.) and 
connections (bridges, jets and filaments). In that case we can add at least 17 
more associations of 3C quasars with low redshift galaxies having chance 
probabilities ranging from $10^{-3}$  to  $10^{-9}$. This seems to already take the 
case for physical association to a very high level of probability. 

\section{Redshift periodicities}

The measured redshift of the QSO, $z = 0.750$, does not lie at one
of the Karlsson peaks.  However, if the galaxy has ejected the
QSO, then its redshift should be calculated relative to the
galaxy. Thus, going back to equation (1) and putting $z_{\rm q} =
0.750$, $z_{\rm g} = 0.344$ we find that

$$(1 + z_{\rm i}) ~ = (1 + z_{\rm q}) / (1 + z_{\rm g}) (1 + z_{\rm d})~] = 1 + 0.302$$

\noindent
which is extremely close to the intrinsic redshift peak at $z_{\rm i}$ = 0.30.

We do not know the value of $z_{\rm d}$, but based on other pairs it has
been shown that $|z_{\rm d}| \leq$ 0.04 (Burbidge and Napier 2001). Thus
the relationship between the observed and predicted values is
very satisfactory. This adds further to the view that the pair is
a true physical system.

\section{Conclusion}

We have discussed this pair of objects from the standpoint of
whether there could be any ``a posteriori quality" to their
extraordinarily small probability of being an accidental
configuration. In fact we have found that this pair has properties
very similar, but more extreme than most of the other associations
of QSOs and galaxies which have been discovered earlier ---
properties of nearness, alignment, disturbances, connections. Since there 
are very few cases that have been examined this closely, the possibility is 
raised that there are more such associations to be discovered. 

We are grateful to Marshall Cohen for sending us details of the optical
spectrum of this object. We thank a referee for a careful examination of the data 
we have discussed, and for emphasizing the need for more optical detail.


\begin{figure}
\includegraphics[width=15.0cm] {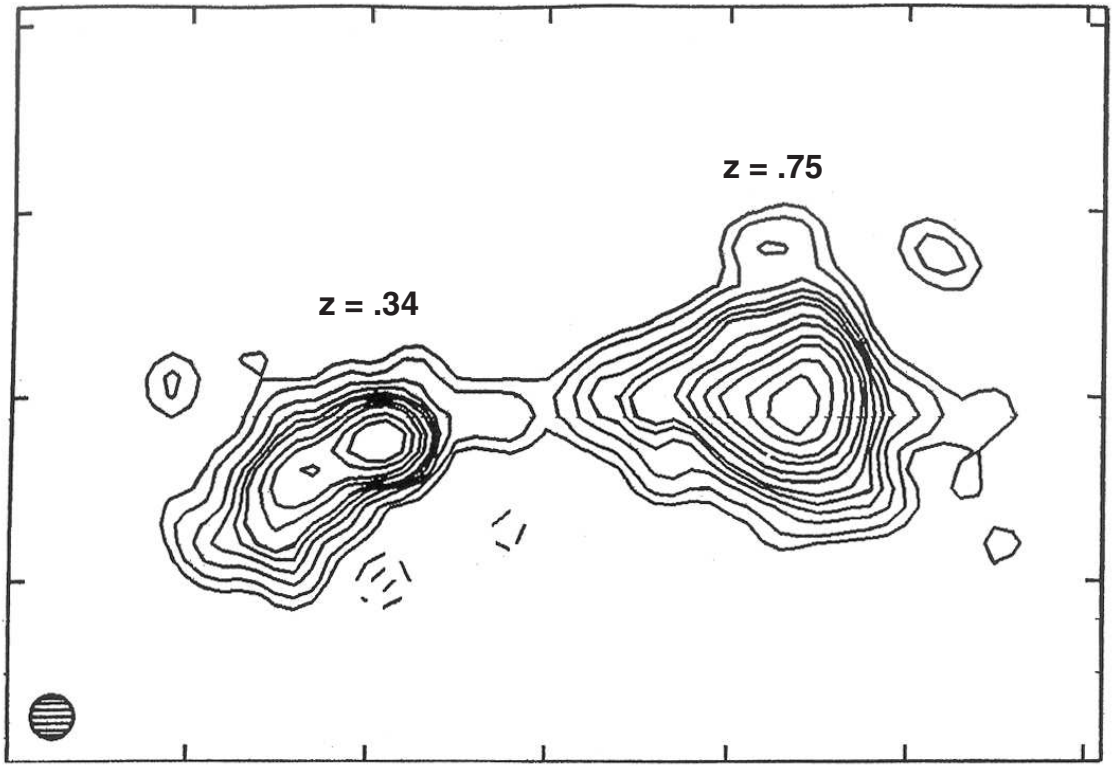}
\caption{Radio map at 1.6 GHz of 3C 343.1 by Fanti et al. (1985).
(The tick marks are separated by $0.1^{\prime\prime}$. The separation of 
sources is about $0.25^{\prime\prime}$). We assign the left hand 
(east) lobe to the galaxy with $z = 0.34$ and the right hand (west) 
lobe to the QSO with $z = 0.75.$
\label{fig1}}
\end{figure}

\end{document}